\documentclass[12pt, twocolumn]{openjournal}
\usepackage{float}
\usepackage{graphicx}
\usepackage{xcolor}
\usepackage{textgreek}
\usepackage[utf8]{inputenc}
\usepackage[english]{babel}
\usepackage{multirow}
\usepackage{hyperref}

\hypersetup{
    colorlinks=true,
    linkcolor=blue,
    filecolor=blue,      
    urlcolor=blue,
    citecolor=blue,
}

\usepackage{color,colortbl}
\usepackage{tensind}
\tensordelimiter{?}
\DeclareGraphicsExtensions{.bmp,.png,.jpg,.pdf}
\usepackage{verbatim}
\usepackage[normalem]{ulem}
\usepackage{orcidlink}
\usepackage{soul}
\usepackage{amsmath}
\usepackage{overpic}

\newcommand{\Ms}{ \mathrm{M}_\odot }

\begin{document}
\title{Planes of satellites, at once transient and persistent}
\author{Till Sawala$^*$ \orcidlink{0000-0003-2403-5358}}
\affiliation{Department of Physics, University of Helsinki, Gustaf H\"allstr\"omin katu 2, FI-00014 Helsinki, Finland}
\email[$^*$Email: ]{till.sawala@helsinki.fi}

\begin{abstract}
The appearance of highly anisotropic planes of satellites around the Milky Way and other galaxies was long considered a challenge to the standard cosmological model. Some recent simulations have found flattened satellite systems to be common, but these have been described as either “transient”, short-lived alignments, or “persistent”, long-lived structures. Here we analyse Milky Way analogue systems in the cosmological simulation TNG-50 to resolve this apparent contradiction. We show that, as the satellite populations of individual hosts rapidly change, the observed spatial anisotropies of their satellite systems are invariably short-lived, with lifetimes of no more than a few hundred million years. However, when the progenitors of the same satellites are traced backwards, we find examples where those identified to form a plane at the present day have retained spatial coherence over several billion years. The two ostensibly conflicting predictions for the lifetimes of satellite planes can be reconciled as two perspectives on the same phenomenon. 
\end{abstract}


\maketitle

\section{Introduction}
\label{sec:intro}
Planes of satellites, or more generally, the anisotropic distributions of satellite galaxies around their hosts, have long been considered a challenge to the $\Lambda$CDM ($\Lambda$ Cold Dark Matter) cosmological model \citep[e.g.][]{Kroupa-2005, Kroupa-2012, Bullock-2017, Perivolaropoulos-2021, Pawlowski-2018, Pawlowski-2021, Boylan-Kolchin-2021, Sales-2022}. In the standard model of hierarchical galaxy and structure formation \citep{Davis-1985}, satellite galaxies populate substructures of only mildly triaxial \citep[e.g.][]{Frenk-1988} dark matter halos. While the distributions of satellite galaxies surrounding the Milky Way \citep[hereafter MW,][]{Lynden-Bell-1976}, M31 \citep{Ibata-2013}, and many other galaxies \citep[e.g.][]{Ibata-2014, Muller-2018}, appear to be highly anisotropic, the conjecture that $\Lambda$CDM predicts fairly isotropic distributions was largely supported by early numerical studies \cite[e.g.][]{Kroupa-2005, Libeskind-2005, Pawlowski-2020}.

More recent, high-resolution simulations have shown that $\Lambda$CDM does predict anisotropic satellite systems, provided artificial numerical disruption of satellites near the centre is avoided \citep[e.g.][]{Sawala-2023a, Xu_2023, Santos-Santos-2023, Zhao-2023, Madhani-2025}. However, the nature of the planes reported by different authors appears incompatible. While some authors have reported that $\Lambda$CDM can only produce short-lived satellite systems that dissolve within a few hundred million years \citep[e.g.][]{Buck-2016, Shao-2019, Santos-2020, Mueller-2021, Samuel_2021, Sawala-2023a, Xu_2023}, others have found long-lived anisotropies that remain intact for several gigayears \citep[e.g.][]{Santos-Santos-2023, Gamez-Martin-2024, Gamez-Martin-2025, Madhani-2025}. 

Furthermore, while \cite{Fernando-2017, Fernando-2018} have shown that planes of satellites within realistic dark matter halos are dynamically unstable and must quickly disperse due to torques imparted in non-spherical potentials and by substructures, others have explained how accretion from the cosmic web \citep[e.g.][]{Libeskind-2015, Gamez-Martin-2024} and/or group infall \citep[e.g.][]{Li-2008, Vasiliev-2024, Taibi-2024, Julio-2024, Callingham-2025} could lead to the emergence of long-lived anisotropies.

With two ostensibly incompatible results for the nature of satellite planes in $\Lambda$CDM now widely discussed, it is unclear what the model actually predicts, and it may even be questioned whether the “plane of satellites problem” is solved at all. In this work, we set out to resolve the apparent inconsistency by showing that, in fact, the same satellite systems can appear either ``transient'' or ``persistent''.

\section{Satellite Systems in TNG-50}
Previous works have used differing definitions of anisotropy or planarity and have analysed different (sometimes unpublished) simulation data. Notwithstanding the merits of more sophisticated metrics, we will focus here on the most widely used measure of spatial anisotropy, $c/a$, where $c$ and $a$ are the square roots of the smallest and largest eigenvalues of the unweighted inertia tensor,
\[
\mathbf{I}_{ij} = \sum_{k=1}^{N} x_{k,i}\, x_{k,j},
\]

where $\mathbf{x}_k$ are the positions of the satellites relative to the host centre. We also limit our presentation to the $N=11$ brightest satellites of each system as analogues to the 11 ``classical'' MW satellites that define its canonical plane, but adopting $N=[9\dots13]$ yields qualitatively similar results.

We consider host and satellite galaxies from the sample of MW analogues described in  \cite{Pillepich-2024}, drawn from the {\sc Illustris TNG-50} cosmological hydrodynamical simulation \citep[hereafter {\sc TNG-50},][]{Nelson-2019a, Nelson-2019b, Pillepich-2019}. The simulation evolves $2160^3$ dark-matter particles with a mass of $4.5\times 10^5\Ms$ and $2160^3$ initial gas cells with an average mass of $8.5\times 10^4\Ms$ in a periodic volume of $51.7^3$~$c$Mpc$^3$ (comoving Mpc), with details provided in the references above.

Planes of satellites in {\sc TNG-50} have already been extensively studied, including by \cite{Pawlowski-2021}, who found a lack of analogues to M31's satellite plane(s), \cite{Kanehisa-2023}, who studied the impact of major mergers on planes of satellites, \cite{Seo-2024}, who reported a low frequency of analogues to the MW's plane of satellites, \cite{Hu-2025}, who found planar structures in ~11\% of MW analogues (although~\cite{Jerjen-2025} show that this result is most likely an artefact of the accidental inclusion of subhalos of non-cosmological origin), \cite{Xu_2023}, who studied in detail a thin but transient plane of satellites around one particular MW analogue, and \cite{Gamez-Martin-2025}, who report the existence of ``kinematically coherent'' subsets of satellites for a number of hosts.

In total, the catalogue of \cite{Pillepich-2024} contains 198 MW analogues at the present time ($z=0$, lookback time $t = 0$). From these, we select the 190 objects (96\%) which are centrals, i.e. the most massive substructures within their respective dark matter halos, so that halo properties such as $\mathrm{r}_{200}$, the radius of a sphere whose density is $200 \times$ the mean density, and $\mathrm{M}_{200}$, the mass enclosed within $\mathrm{r}_{200}$, can be identified and directly attributed to them.

\begin{figure*}

\centering 
\begin{overpic}[width=15.6cm, trim={0.2cm 0.2cm 0.2cm 0.cm},clip]{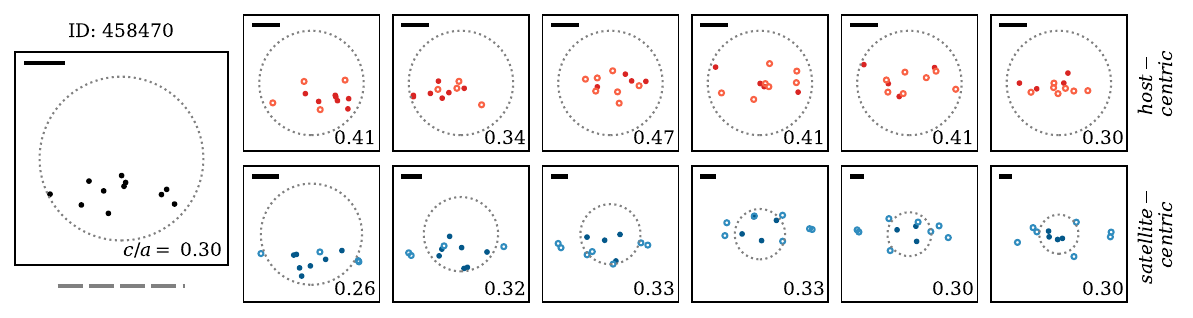}
\put(8,27) {$t=0$}
\put(24.5,27) {$t=0.5$}
\put(38,27) {$t=1.0$}
\put(51.5,27) {$t=1.5$}
\put(64,27) {$t=2.0$}
\put(77.5,27) {$t=2.5$}
\put(91,27) {$t=3.0$}

\end{overpic}

\vspace{.25cm}
\includegraphics[width=15.6cm, trim={0.2cm 0.2cm 0.2cm 0.2cm},clip]{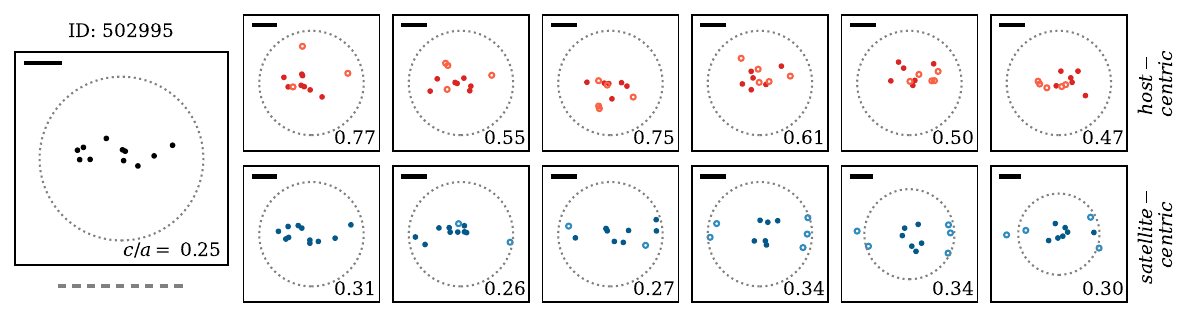}\\
\vspace{.25cm}
\includegraphics[width=15.6cm, trim={0.2cm 0.2cm 0.2cm 0.2cm},clip]{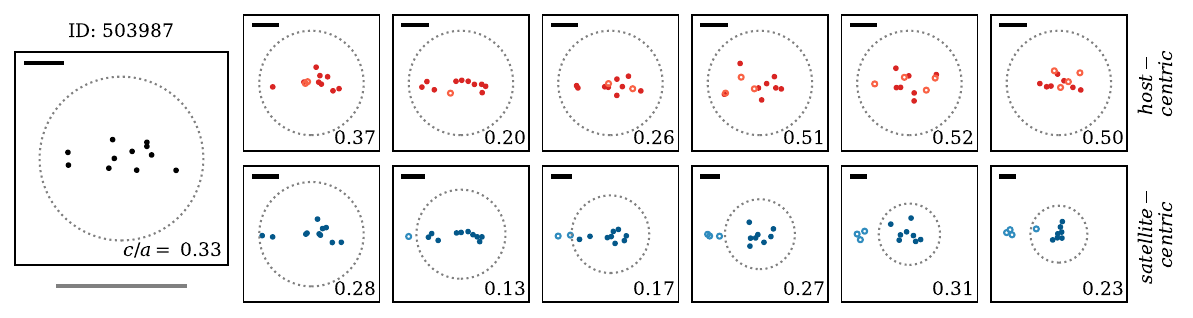}\\
\vspace{.25cm}
\includegraphics[width=15.6cm, trim={0.2cm 0.2cm 0.2cm 0.2cm},clip]{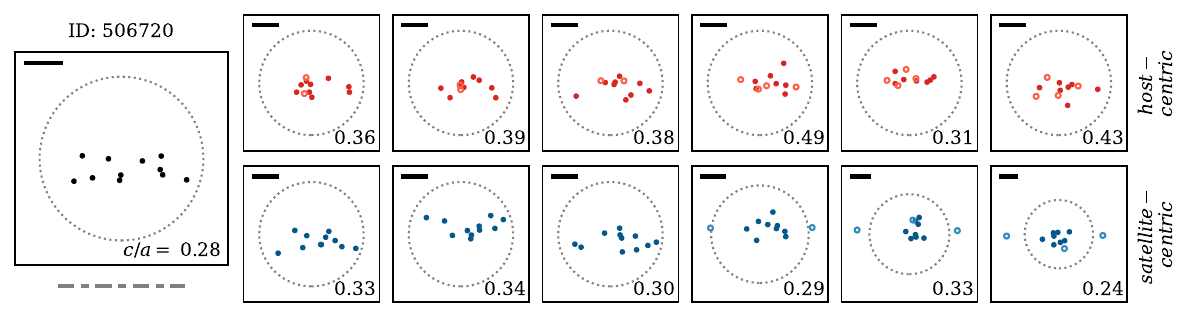}\\
\vspace{.25cm}

\caption{\label{fig:positions} Position of the 11 brightest satellites, projected edge-on (i.e. in the plane of the minor and major axes of the inertia tensor), in four example systems, at the present day (left), and six lookback times up to 3 Gyr, in the host-centric frame (top, red) and the satellite-centric frame (bottom, blue). In the host-centric frame, filled circles denote satellites that are part of the $t=0$ sample, in the satellite-centric frame, filled circles denote subhalos inside $r_\mathrm{lim}(t)$. On each panel, a dotted circle of radius $r_\mathrm{lim}(t)$ and a solid bar of length 100~kpc are shown for scale, and the value of $c/a$ is given. The ID above the $t=0$ panel identifies the host subhalo in the simulation at $t=0$, and the line style below the $t=0$ panel identifies the system in Figure~\ref{fig:ca-evolution}. In these four systems, the spatial anisotropy varies rapidly in the host-centric frame, but remains stable in the satellite-centric frame.
}\vspace{.15cm}
\end{figure*}

\begin{figure*}
\centering 
\includegraphics[width=8.9cm, trim={0.7cm 0.2cm 0.0cm 0.0cm},clip]{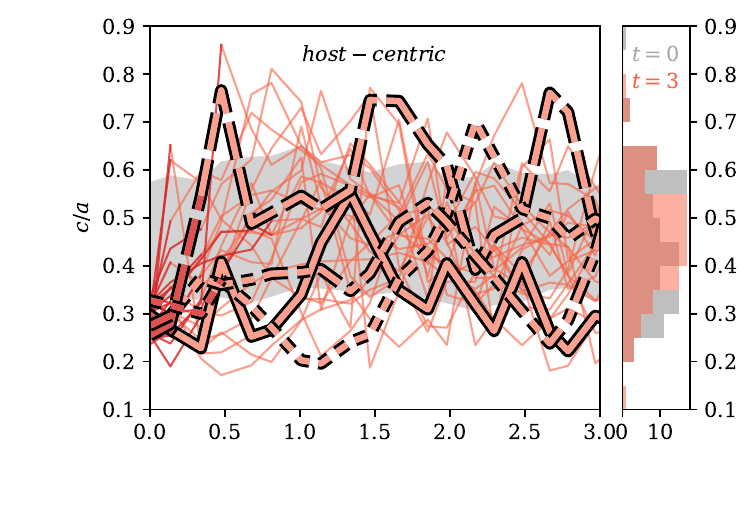}
\includegraphics[width=8.9cm, trim={0.7cm 0.2cm 0.0cm 0.0cm},clip]{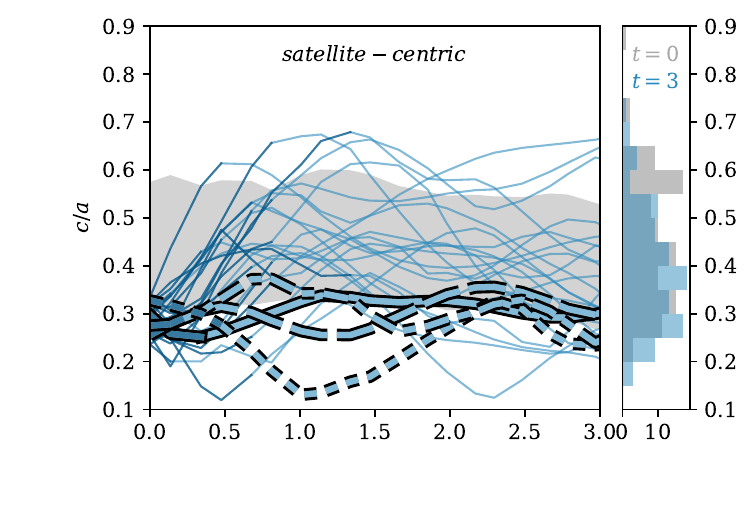} \\ \vspace{-.5cm}
\includegraphics[width=8.9cm, trim={0.7cm 0.2cm 0.0cm 0.0cm},clip]{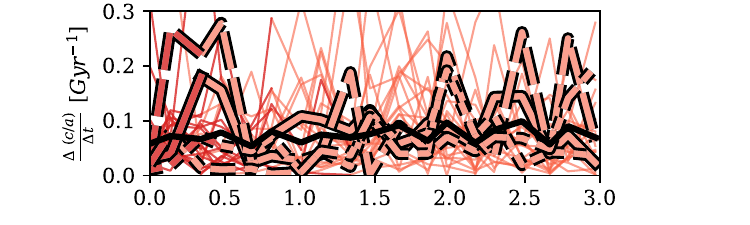} 
\includegraphics[width=8.9cm, trim={0.7cm 0.2cm 0.0cm 0.0cm},clip]{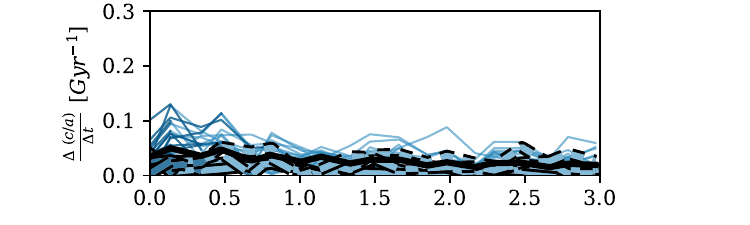} \\ \vspace{.15cm}
\includegraphics[width=8.9cm, trim={0.7cm 0.2cm 0.0cm 0.0cm},clip]{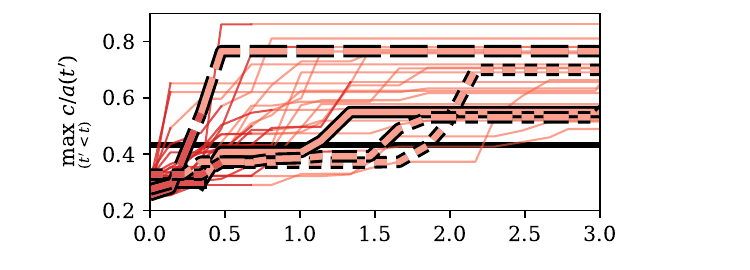} 
\includegraphics[width=8.9cm, trim={0.7cm 0.2cm 0.0cm 0.0cm},clip]{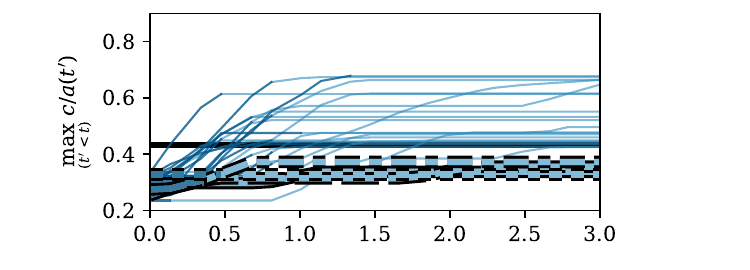} \\ \vspace{.15cm}
\includegraphics[width=8.9cm, trim={0.7cm 0.cm 0.0cm 0.0cm},clip]{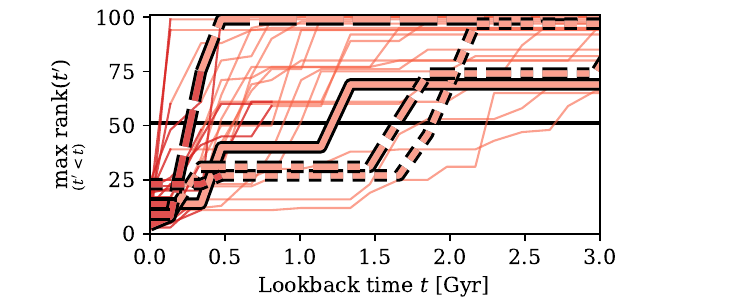} 
\includegraphics[width=8.9cm, trim={0.7cm 0.cm 0.0cm 0.0cm},clip]{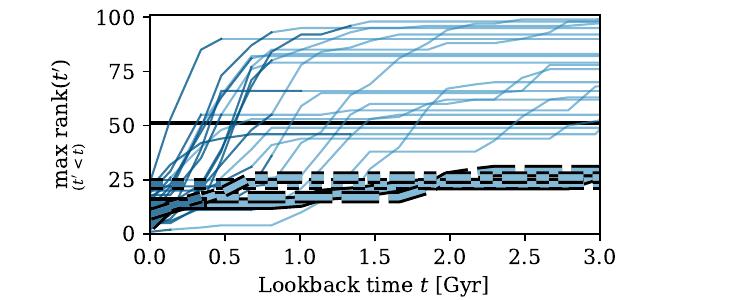}
\vspace{.15cm}
\caption{\label{fig:ca-evolution} Top row: evolution of the anisotropy, $c/a$, in the host-centric frame (left) or satellite-centric frame (right). Second row: rate of change of $c/a$ between individual snapshots. Third row: maximum value of $c/a$ up to lookback time $t$. Fourth row: maximum rank, by $c/a$, up to lookback time $t$. Individual lines show the 25 systems in the lowest quartile of $c/a$ at $t=0$, with the four systems described in Figure~\ref{fig:positions} highlighted. Dark line segments indicate the interval in which the system contains the same satellites as at $t=0$ (host-centric frame), or when all satellites are within $r_{\mathrm{lim}}$ (satellite-centric frame). Grey bands on the top panels show $\pm 1 \sigma$-equivalent percentiles for all 101 systems, with histograms showing the corresponding distributions at $t=0$ and $t=3$~Gyr. Black lines in the second row show the moving median of $c/a$, black lines in the third and bottom row show the median at $t=0$ and the median rank, respectively. Only in the satellite-centric frame do several systems maintain a persistently high anisotropy.}
\vspace{.15cm}
\end{figure*}

We use the location of the most bound particle of a subhalo to define its position, and identify as satellite galaxies all subhalos that contain stars, and which are located within $r_\mathrm{lim}(t) = 1.2 \times \mathrm{r}_{200}(t)$ from the host. This factor results in a median of $r_\mathrm{lim}(t=0) = 277$~kpc and corresponds to the canonical definition of $r_\mathrm{lim}(t=0) = 300$~kpc for studies of MW satellites, whose $\mathrm{r}_{200}$ at $t=0$ is estimated to be $\sim 250$~kpc (choosing $r_\mathrm{lim} = \mathrm{r}_{200}$ slightly reduces the sample size and the typical time satellites spend inside $r_\mathrm{lim}$, but yields very similar results for all metrics we discuss). By this definition, 188 out of 190 systems (99\%) have at least 11 satellite galaxies at $t=0$. Where there are more than 11 satellites, we choose the subset with the highest stellar masses.

Additionally, we require that the host and all 11 satellites identified at $t=0$ are genuine subhalos of cosmological origin ( \texttt{SubhaloFlag == True}), have main-branch progenitors for at least 19 additional outputs (lookback time of 3.27 Gyr, corresponding to a redshift of $z=0.28$ and a mean snapshot interval of $\sim 0.16 $~Gyr), a condition fulfilled by 109 systems (57\%), and that at least 11 satellite galaxies are found within $r_\mathrm{lim}(t)$ of the particular host for the same time interval, which 181 systems (95\%) satisfy. This choice of lookback time interval represents a compromise: at $t>3 Gyr$, we can clearly distinguish ``short-lived'' and ``long-lived'' features, while also retaining a large enough sample of systems for which we can trace the progenitors of all $N=11$ satellites, given the finite mass resolution of the simulation. Due to the finite time interval, some satellites which we identify as newly accreted may have had earlier pericentre passages, and some degree of coherence due to subhalos with very long orbital periods could be missed for this reason.

We remove one system whose $\mathrm{r}_{200}$ repeatedly changes by more than $50\%$ between adjacent snapshots in our lookback time interval, indicating a close interaction with a more massive halo and resulting in discontinuities in our satellite definition. None of the remaining systems change in $\mathrm{r}_{200}$ by more than $10\%$ between any two snapshots.

Our final sample contains 101 of the 190 systems (53\%) that fulfil all criteria. While our selection could introduce biases (such as preferentially selecting higher-mass systems), the halo mass distribution closely matches the parent sample ($M_{200,N=190} = 1.36^{+1.13}_{-0.53} \times 10^{12}\Ms$ vs. $M_{200,N=101} = 1.30^{+1.08}_{-0.43} \times 10^{12}\Ms$), and is compatible with mass estimates of $\sim 10^{12} \Ms$ for the MW \citep[e.g.][]{Sawala-2023b, Hunt-2025}.

It must be noted that, despite its relatively high mass resolution, TNG-50 likely still suffers from the artificial disruption of satellites \citep{vandenBosch-2018}. For comparison, \cite{Grand-2021} showed that an even higher mass resolution of $3\times 10^5 \Ms$ is still insufficient to prevent numerical disruption of substructures near the centre. As shown by \cite{Sawala-2023a}, one consequence of this is a much lower detectable anisotropy, making simulations like {\sc TNG-50} unsuitable for quantifying the frequency of satellite planes predicted by the underlying cosmological model. In our case, none of the systems in the sample reaches the MW’s low $c/a = 0.183$, and only one matches the MW’s high Gini coefficient of inertia \citep[see][]{Sawala-2023a}, an indicator of artificial disruption and artificially reduced anisotropy.

However, we are not concerned with quantifying the incidence of satellite planes, but in their time evolution. For this purpose, we consider systems with $c/a$ in the lowest quartile ($c/a < 0.34$) at $t=0$. The large sample size, the accessibility of the data, and the extensive prior analysis make {\sc TNG-50} ideal for demonstrating a more general principle.

\section{Host-centric and satellite-centric frames}
We follow the evolution of the satellite systems and their spatial anisotropies in two complementary frames. In the host-centric frame, we follow the main progenitor of each MW analogue backwards in time. At each output, we re-select all subhalos within $r_\mathrm{lim}(t)$ of the host's centre at that time, and measure the properties of the 11 satellites with the greatest stellar mass. In the satellite-centric frame, we follow the main progenitors of each of the 11 most massive satellites identified at $t=0$ over the same time interval. At each output, we measure $c/a$ for the same subhalos, irrespective of their positions and their stellar-mass rank at that time.

In the host-centric frame, we thus trace the evolution of the anisotropy of the (potentially changing) satellite population surrounding each host. In particular, we may ask how long a given satellite plane found at $t=0$ would have been identified surrounding the same galaxy, had its satellite system been examined independently and under the same conditions under which it was examined at $t=0$. Additionally, we track the change in the makeup of the satellite population, and ask for how long the {\it same} subset of subhalos would have been identified as each host's 11 brightest satellites.

In the satellite-centric frame, we always follow the evolution of a fixed set of subhalos identified to be the 11 most massive satellites of the host at $t=0$, irrespective of whether they were among the 11 brightest satellites or whether they were even within $r_{lim}$ at some earlier time. In this way, we may ask for how long the particular subset of subhalos identified as an anisotropic satellite system at $t=0$ had maintained its anisotropy.

Both perspectives are identical at $t=0$, but diverge with increasing lookback time, as new satellites have fallen in, and previously existing ones have become disrupted. As we will show, in the host-centric frame, any host's satellite plane has been short-lived, while the satellite-centric frame reveals several examples of long-lived anisotropies.

\vspace{1cm}

\begin{figure*}
\centering 

\includegraphics[width=8.6cm, trim={0.3cm 0.2cm 0.0cm 0.0cm},clip]{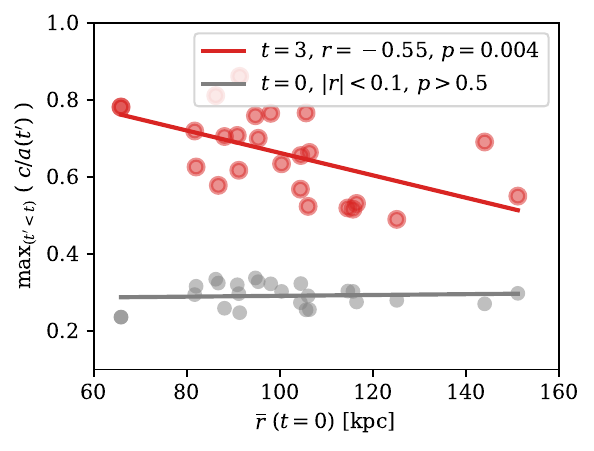}
\includegraphics[width=8.6cm, trim={0.3cm 0.2cm 0.0cm 0.0cm},clip]{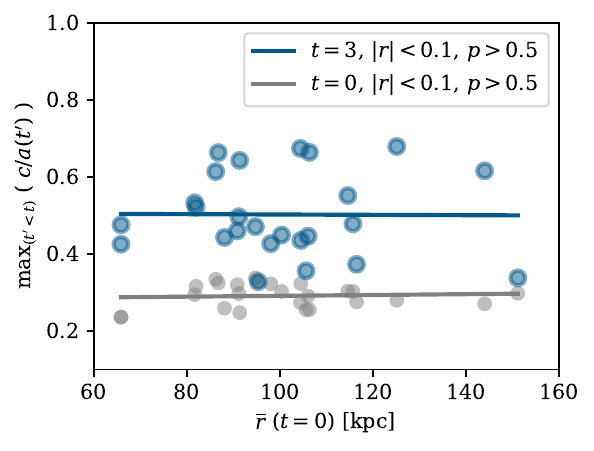}

\caption{Dependence of the anisotropy evolution on the mean Galactocentric distance, $\overline{r}$, in the host-centric frame (left) and the satellite-centric frame (right). Grey symbols show $c/a$ as a function of $\overline{r}$ at $t=0$, coloured symbols show the maximum value of $c/a$ during 3 Gyr of lookback time. Lines show linear least-square fits, $r$ denotes the Pearson correlation coefficient, $p$ denotes the two-sided probability of obtaining a correlation of equal or greater magnitude from a sample with no underlying correlation. At $t=0$, there is no significant correlation between $r$ and $c/a$. In the host-centric frame, more extended satellite systems evolve more slowly; in the satellite-centric frame, there is no significant correlation.} \label{fig:r-dependence}
\vspace{.3cm}
\end{figure*}

\begin{figure*}
\centering 

\includegraphics[width=8.6cm, trim={0.3cm 0.72cm 0.0cm 0.0cm},clip]{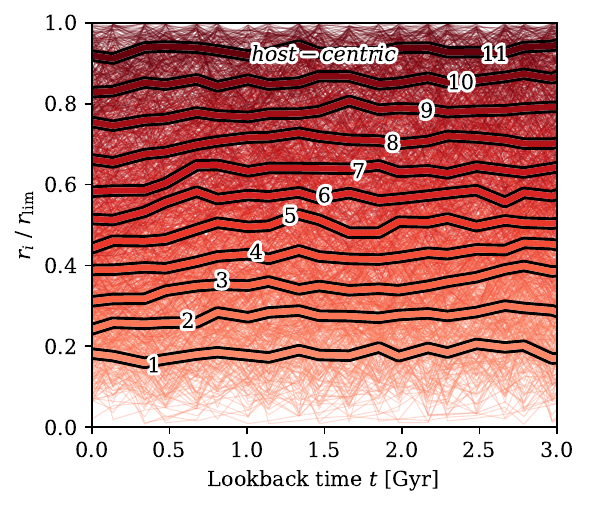}
\includegraphics[width=8.6cm, trim={0.3cm 0.72cm 0.0cm 0.0cm},clip]{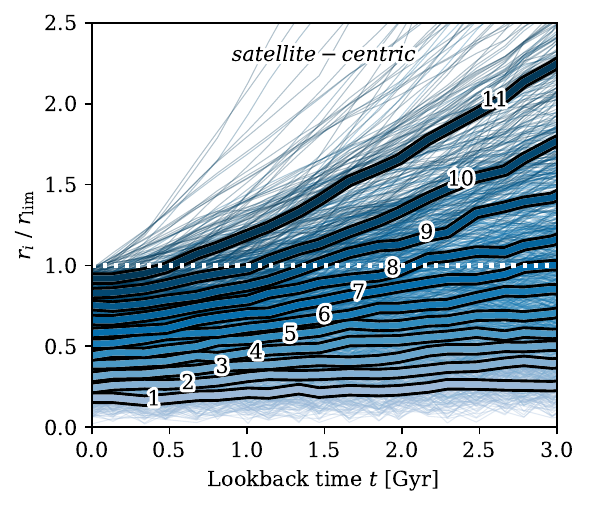}
\includegraphics[width=8.6cm, trim={0.3cm 0.2cm 0.0cm 0.0cm},clip]{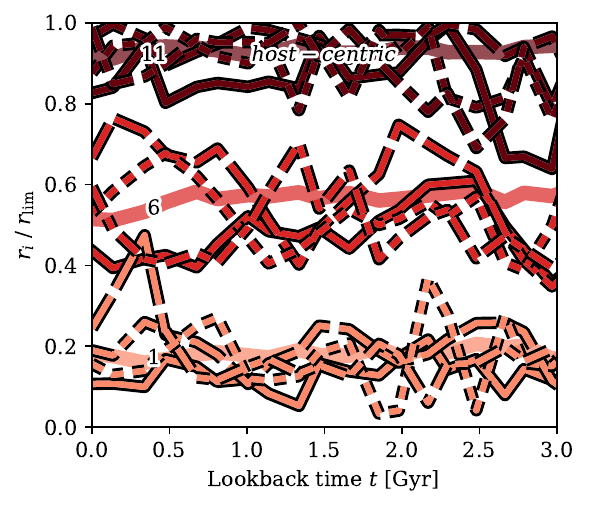}
\includegraphics[width=8.6cm, trim={0.3cm 0.2cm 0.0cm 0.0cm},clip]{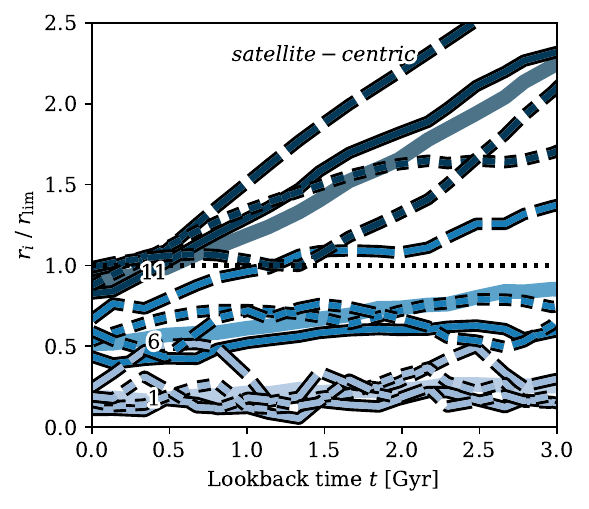}

\caption{\label{fig:radii} Evolution of the distance of the $i_\mathrm{th}$ innermost satellite to the centre, normalised by $r_\mathrm{lim}(t)$ , in the host-centric frame (left), or  satellite-centric frame (right). In the top panels, thick lines show the median ratio, thin lines of corresponding colours show individual systems. In the bottom panels, individual lines show the corresponding radii of the $1^{st}$ (innermost), $6^{th}$ (median), and $11^{th}$ (outermost satellite), broad bands replicate the corresponding medians from the top panels. In the host-centric frame, by definition, all satellites are always inside $r_\mathrm{lim}(t)$, and normalised by $r_\mathrm{lim}(t)$, the average radial profile shows little evolution. In the satellite-centric frame, the system ``expands'' with increasing lookback time. Of the 11 satellites at $t=0$, two typically fell in less than one Gyr ago, and four less than two Gyr ago. The evolution of the highlighted systems is typical of the sample in this perspective.}
\vspace{.3cm}
\end{figure*}

\section{The anisotropy evolution}
Figure~\ref{fig:positions} shows four satellite systems that are anisotropic at $t=0$. For each system, the larger, leftmost panels show the positions of the 11 brightest satellites at $t=0$. The next six rows show the positions at lookback times up to 3 Gyr, in both the host-centric frame (upper row, red) and in the satellite-centric frame (lower row, blue). On all panels, $r_\mathrm{lim}(t)$ of the host halo is shown for scale.

These examples show clear differences. Firstly, in the host-centric frame, the identity of the 11 brightest satellites changes quickly: in all four cases, at least two satellites are different at a lookback time of only 0.5 Gyr. This change in the composition of satellite samples is caused by two effects: satellites that are part of the $t=0$ sample but which have only recently fallen in, and satellites that were part of the earlier sample, but which have been tidally stripped or destroyed at $t=0$. Both contribute to the rapid evolution of the anisotropy in the host-centric frame.

By contrast, in the satellite-centric frame, at earlier times, an increasing fraction of the satellites that define the plane at $t=0$ are found outside $r_\mathrm{lim}$. In particular, at larger lookback times, the progenitors of present-day plane members can be far outside their (future) host's halo. Subhalos outside the halo are not subject to the torques that change their orbits within the halo. The result is a much more gradual evolution of the anisotropy, and the existence of long-lived anisotropies previously reported.

This stark difference in evolution between the two frames can also be observed in larger samples. The top row of Figure~\ref{fig:ca-evolution} shows the evolution of the anisotropy for all systems with $c/a$ in the lowest quartile ($c/a < 0.34$) at $t=0$. The second row shows the absolute rate of change in anisotropy between adjacent snapshots, $\Delta(c/a)/\Delta t$. The third and fourth row show the running maximum of $c/a$ as a function of lookback time, $\max_{t'<t}\left(c/a\left(t'\right)\right)$, and the maximum rank by $c/a$. For panels on the left, the host-centric frame is adopted, while those on the right show the satellite-centric view. Each line represents an individual system, bold lines identify the four systems shown in Figure~\ref{fig:positions}. 

In the host-centric frame, for each system, dark line segments denote the time intervals when the satellite sample consists of the same 11 individual satellites identified at $t=0$, while light line segments indicate a change in the satellite composition. In the satellite-centric frame, dark segments denote the time intervals when all 11 subhalos are within $r_\mathrm{lim}$. None of the 25 systems maintain the same 11 brightest satellites for more than 1~Gyr in the host-centric frame, and in no case do all the progenitors of the 11 brightest satellites remain within $r_\mathrm{lim}$ for more than 1.5~Gyr.

Starting from the same high anisotropies at $t = 0$, most systems show a ``regression to the mean'' in both the host-centric frame and the satellite-centric frame, consistent with chance alignments in both frames. However, the anisotropy changes much more rapidly, and more universally, in the host-centric frame. Here, even the slowest-changing systems undergo rapid changes in anisotropy, and no system has maintained a value of $c/a$ below the median value, or a rank below the median rank, throughout the 3 Gyr time interval. In the host-centric frame, all satellite planes would be considered short-lived.

While large (although not as rapid) changes are also common in the satellite-centric frame, several systems, including those highlighted, only undergo a mild change in anisotropy, maintaining close to their present value for several Gyr in lookback time, and consistently ranking among the most anisotropic systems. In the satellite-centric frame, several systems have maintained long-lived anisotropies, but these would be detected as ``planes of satellites'' for only a fraction of the time, as some of the subhalos that constitute the $t=0$ planes have only recently been accreted.

Figure ~\ref{fig:r-dependence} shows the present-day anisotropy, $c/a$, and the maximum value attained during 3 Gyr, $\mathrm{max}(c/a)$, as functions of the average distance of satellites from the host centre, $\overline{r}$, for the same systems in the lowest quartile of $c/a(t=0)$ identified in Figure~\ref{fig:ca-evolution}. At $t=0$, where both frames are identical,  there is no significant correlation between $r$ and $c/a$. In the host-centric frame, the radial distribution of satellites affects the rate at which the anisotropy dissolves: systems with more extended satellites evolve more slowly than those with more compact satellite distributions. In the satellite-centric frame, there is no significant correlation between the present-day radial distribution and the change in anisotropy.

Several factors contribute to changes in the satellite population in the host-centric frame with increasing lookback time: newly infalling bright satellites, satellites becoming disrupted or merging with the host halo, and satellites exiting the boundary of their hosts (so-called ``backsplash'' galaxies).  In total, 2685 distinct galaxies (i.e. without duplicating those that are the descendants of satellites from earlier outputs) have been among the $N=11$ brightest satellites across the 101 systems and across the 20 outputs analysed. Of those, $101 \times 11 = 1111$ $(\sim41\%)$ are the present-day brightest satellites. Among the remaining ones, 742 $(\sim28\%)$ have subsequently merged with the main halo, 457 $(\sim17\%)$ have remained satellites but no longer rank among the brightest 11 at $t=0$, and 375 $(\sim14 \%)$ have become ``backsplash'' galaxies, located outside $r_\mathrm{lim}$ at $t=0$.

Considering only those 1111 subhalos that were the brightest satellites at a lookback time of 3~Gyr, we find that 608 $(\sim55\%)$ are still among the brightest satellites at $t=0$, 112 $(\sim10\%)$ have merged with the main halo, 208 $(\sim19\%)$ remain satellites, but no longer rank among the brightest 11, and 183 $(\sim16\%)$ have become ``backsplash'' galaxies.

In the satellite-centric frame, the systems effectively expand beyond the host halo with increasing lookback time, maintaining or often increasing their anisotropies. By contrast, in the host-centric frame, the system approximately retains its size, but the subhalo populations are exchanged over time.

It is worth noting that observations of Milky Way satellites support a picture where several satellites have fallen in relatively recently. In particular, using Gaia data, \cite{Taibi-2024} show that members of the satellite plane are predominantly late-infalling satellites.

\begin{figure}
\centering 
\includegraphics[width=8.6cm, trim={0.2cm 1.2cm 0.2cm 0.0cm},clip]{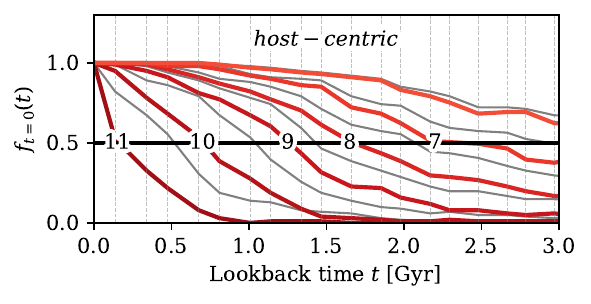}
\includegraphics[width=8.6cm, trim={0.2cm 0.2cm 0.2cm 0.0cm},clip]{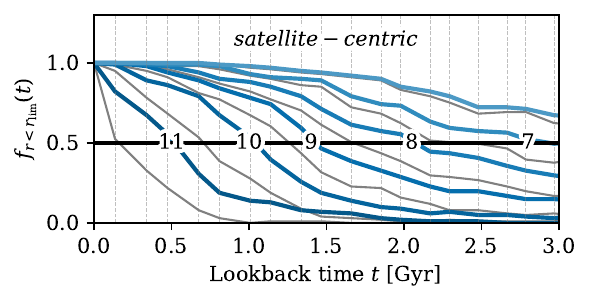}

\caption{\label{fig:fractions} Top: Fraction of systems for which at least $N$ of 11 satellites of the $t=0$ system are part of the system in the host-centric frame. Bottom: Fraction of systems for which at least $N$ of 11 satellites are found within $r_\mathrm{lim}$ in the satellite-centric frame. Grey lines show the corresponding lines of the other panel, vertical lines indicate the timing of snapshots.}
\vspace{.15cm}
\end{figure}

\begin{figure}
\centering 
\includegraphics[width=8.6cm, trim={0.05cm 1.2cm 0.0cm 0.0cm},clip]{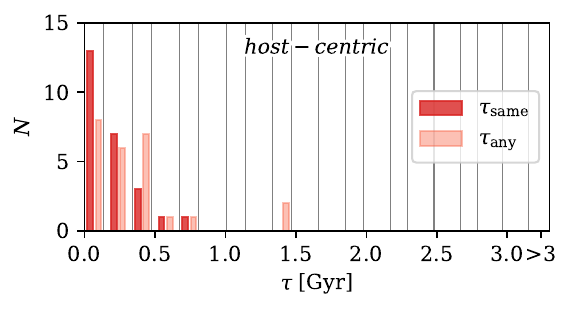}
\includegraphics[width=8.6cm, trim={0.05cm 0.2cm 0.0cm 0.0cm},clip]{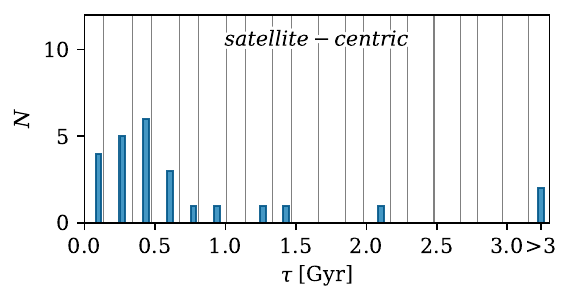}

\caption{\label{fig:lifetimes} Lifetime of satellite planes in the host-centric frame (top) and the satellite-centric frame (bottom). In the host-centric frame, we distinguish between planes that contain the {\it same} satellites, and {\it any} satellite plane in the same halo (including those whose membership has changed). Vertical lines indicate the timing of the snapshots.}\vspace{.15cm}
\end{figure}

In Figure~\ref{fig:radii}, the difference is illustrated by considering the radial distribution of satellites in both the host-centric and the satellite-centric frame. In the host-centric frame, where satellites are identified anew within $r_\mathrm{lim}$ at each snapshot, the average radial distribution does not change, with the innermost of the 11 satellites typically found close to $0.2 \times r_\mathrm{lim}$, and the outermost typically close to $0.9 \times r_\mathrm{lim}$. By contrast, in the satellite-centric frame, where individual satellites are traced, the systems ``expand'' with increasing lookback time: one satellite is typically outside $r_\mathrm{lim}$ in under 0.5 Gyr, and by 3 Gyr, nearly half of the subhalos identified at $t=0$ had not yet fallen into $r_\mathrm{lim}$.

Figure~\ref{fig:fractions} shows the fraction of systems, as a function of time, for which the $N$ satellites are within $r_\mathrm{lim}$ in the satellite-centric frame, and the fraction of systems, as a function of time, for which the set of satellites in the host-centric frame contains $N$ of the same satellites as those identified at $t=0$.

As subhalos cross $r_\mathrm{lim}$, both fractions decrease with increasing lookback time. In the satellite-centric frame, the sets of subhalos contain an increasing number of subhalos outside $r_\mathrm{lim}$. In the host-centric frame, these subhalos are replaced by different subhalos within $r_\mathrm{lim}$. However, for a given $N$, the decrease in $f_{t=0}$ proceeds slightly faster than that in $f_{r< r_\mathrm{lim}}$. This is because the replacement of subhalos can also be caused by the appearance, at increasing lookback times, of satellites that were subsequently destroyed (or in rare cases, escaped) by $t=0$. \\

\section{Lifetimes of satellite planes}
In Figure~\ref{fig:lifetimes}, we show the distribution for the lifetime, $\tau$, of satellite planes, i.e. of satellite systems in the lowest quartile of $c/a$ at $t=0$. In the host-centric frame, we define $\tau_\mathrm{same}$ as the time for which the {\it same} set of satellites identified at $t=0$ has maintained a value of $c/a$ in the lowest quartile within $r_\mathrm{lim}$, and $\tau_\mathrm{any}$ as the time for which the 11 brightest satellites of the system have maintained the same anisotropy, regardless of membership changes over time. In the satellite-centric frame, $\tau$ is the time for which the same set of subhalos has maintained a value in the lowest quartile, whether inside or outside of the halo.

Both due to the dispersal of satellite positions, and due to the changing satellite composition, satellite planes in the host-centric frame have short lifetimes. The only examples of satellite planes with lifetimes beyond 0.6 Gyr in the host-centric frame are those in which the composition of satellites has changed over time. By contrast, in the satellite-centric frame, several systems maintain planes for at least 3 Gyr. In all of these cases, the anisotropy has invariably been maintained outside of the host halo, and before some of the subhalos constituting the $t=0$ plane have become satellites.

\section{Conclusion}
Simulations in the $\Lambda$CDM paradigm can produce highly anisotropic satellite systems, but their evolution and apparent lifetime is a matter of perspective. Within any given halo, planes of satellites are transient, identifiable for only a few hundred Myr. However, the progenitors of the present-day satellites may have maintained persistent coherence for several Gyr, beyond the time that many of them have been satellites of the same halo.

Differences between previous studies also arise from applying different anisotropy metrics to different subsets of satellites in different simulations. For example, \cite{Santos-Santos-2023} and \cite{Gamez-Martin-2025} have pointed out that subsets of satellites with aligned orbital poles can exhibit persistent spatial anisotropies. However, our results suggest that the disagreement on whether spatial anisotropies are persistent or transient may arise at an even more basic level from the choice of reference frame. Studies that have found only short-lived features have largely considered the evolution of halos and their evolving satellite systems in the host-centric frame \citep{Shao-2019, Samuel_2021, Sawala-2023a}. By contrast, those studies that have found long-lived features have invariably adopted the satellite-centric frame \citep{Santos-Santos-2023, Gamez-Martin-2024, Gamez-Martin-2025}, tracing the history of the satellite progenitors themselves, often beyond their infall times. Notably, \cite{Bahl-2014} and \cite{Xu_2023} also adopted the satellite-centric frame, but reported only short-lived planes, illustrating that this frame may be a necessary, but certainly not sufficient condition for detecting longevity.

This understanding also reconciles the two apparently conflicting theoretical predictions: the anisotropic and correlated distributions of satellites before infall predicted by $\Lambda$CDM can lead to anisotropic satellite systems, which, when traced back in time, appear long-lived \citep[e.g.][]{Li-2008, Libeskind-2015, Gamez-Martin-2024}. However, this is not in conflict with the finding that, once within a halo, satellite planes must be short-lived \citep{Fernando-2017, Fernando-2018}. The former considers anisotropies in the satellite-centric frame, while the latter considers the host-centric frame. The longevity of planes in the satellite-centric frame relies on the fact that satellites only spend a short time inside the halo, where any planes quickly dissolve.

It is worth repeating that, most likely due to resolution effects, none of the systems in {\sc TNG-50} are as flattened as the Milky Way's plane of satellites. The resolution may also affect the time evolution of satellite planes in {\sc TNG-50}, but this is unlikely to be the main driver for the different behaviour between the two frames we describe here. Future, higher resolution simulations will be required to confirm this, and to extend our analysis to planes with genuine high anisotropies.

Our work cannot directly address the question of whether the actual Milky Way's plane of satellites is transient. Simple backwards orbital integration suggests that it disperses within a few hundred Myr of lookback time, even under idealised conditions \citep[e.g.][]{Maji-2017, Sawala-2023a}, although \cite{Kumar-2025} caution that this may underestimate the effect of observational errors. Perhaps more importantly, by integrating closed orbits of the present-day satellites in a static potential, these studies neither account for the past disruption of satellites, nor for the finite infall times of the present satellite population. Our results suggest that both of these effects could potentially shorten the lifetime of the MW's plane in a host-centric frame. A clear answer and a more meaningful comparison to simulation results may therefore not only require more precise observations of the present-day satellite system, but also accounting for the evolving potential~\citep{Martinez-Garcia-2025}.

We conclude that, as a prediction of the $\Lambda$CDM model, within a given galaxy, satellite planes represent only a brief phase, with lifetimes that rarely exceed a few hundred million years. However, these satellite planes are not all ephemeral, chance alignments: they can consist of substructures whose anisotropies can be traced back over several Gyr. Satellite planes that are transient in the host-centric frame can be persistent in the satellite-centric frame.

\section*{Code \& Data Availability}
Documented code to reproduce all results and figures presented in this paper is provided at: \\ \url{www.github.com/TillSawala/transient-persistent}. \\ Reduced data and variants of the plots for $N=[9\dots13]$ satellites and for $r_\mathrm{lim} = \mathrm{r}_{200}$ are included. The code can also easily be used to consider, for example, different numbers or subsets of satellites, different definitions of $r_\mathrm{lim}$, of the lifetime, $\tau$, or different lookback time intervals. The underlying simulation data is available at: \url{www.tng-project.org}.

\section*{Acknowledgements}
I thank Isabel Santos-Santos, Jenni Häkkinen, Peter Johansson and Gabor Racz for very helpful comments, and I thank the reviewers for their helpful and constructive comments and suggestions. I thank the creators of the {\sc TNG-50} simulation for making their data public, and the Flatiron Institute and the Simons Foundation for travel support at the start of the project. I acknowledge support by Research Council of Finland grants 354905 and 339127, and ERC Consolidator Grant KETJU (no. 818930). This work used facilities hosted by the CSC—IT Centre for Science, Finland. I also gratefully acknowledge the use of open-source software, including \texttt{Matplotlib} \citep{matplotlib-paper}, \texttt{SciPy} \citep{SciPy} and \texttt{NumPy} \citep{numpy-paper}.

\bibliographystyle{mnras}
\bibliography{bibliography}
\vfill\null\vspace{2.3cm}\vfill\null
\end{document}